\documentclass[twocolumn,epjc3]{svjour3}
\smartqed  
\RequirePackage{graphicx}
\usepackage{amsmath,amsfonts,euscript,amssymb}
\usepackage[T1]{fontenc}
\usepackage[english]{babel}
\usepackage{url}

 \newcommand{\be}{\begin{equation}}
 \newcommand{\ee}{\end{equation}}
 \newcommand{\bea}{\begin{eqnarray}}
 \newcommand{\eea}{\end{eqnarray}}

\DeclareSymbolFont{bfitletters}{OML}{cmm}{bx}{it}
\DeclareSymbolFont{bfitoperators}   {OT1}{cmr} {m}{n}
\DeclareMathSymbol{\bfitomega}{\mathord}{bfitletters}{"21}
\DeclareMathSymbol{\bfitrho}{\mathord}{bfitletters}{"1A}
\DeclareMathSymbol{\bfitgamma}{\mathord}{bfitletters}{"0D}
\DeclareMathSymbol{\bfitchi}{\mathord}{bfitletters}{"1F}
\DeclareMathSymbol{\bfitxi}{\mathord}{bfitletters}{"18}

\journalname{}
\begin{document}

\title{Reduced Geometrodynamics of Closed Manifolds}



\author{Andrej B. Arbuzov\thanksref{e1,addr1,addr2}
        \and
        Alexander E. Pavlov\thanksref{e2,addr3}
}

\thankstext{e1}{e-mail: arbuzov@theor.jinr.ru}
\thankstext{e2}{e-mail: alexpavlov60@mail.ru}


\institute{Bogoliubov Laboratory of Theoretical Physics,
           Joint Institute for Nuclear Research,
           Joliot-Curie str. 6, Dubna, 141980, Russia \label{addr1}
        \and
           Department of Higher Mathematics, Dubna State University,
           Universitetskaya str. 19, Dubna, 141980, Russia \label{addr2}
        \and
          Institute of Mechanics and Energetics, Russian State Agrarian University,
          Timiryazevskaya str., 49, Moscow 127550, Russia \label{addr3}
}

\date{\today}

\maketitle

\begin{abstract}
The global time is defined in covariant form under the condition of
a constant mean curvature slicing of spacetime. The background static
metric is taken in the tangent space. The global intrinsic time is
identified with the logarithmic function of the mean value of
the ratio of the square root of the metric determinants. The procedures
of the Hamiltonian reduction and deparametrization of dynamical
systems are implemented. The Hamiltonian system appears to be non-conservative.
The Hamiltonian equations of motion of gravitational field in the global time
are written.
\end{abstract}

\section{Introduction}
\label{intro}

The problem of global time is basic in the General Relativity~\cite{Kuchar,Isham}.
The conservative energy is well defined in asymptotically flat spaces because of
the existence of the preferable coordinate frame with a coordinate time~\cite{Arnowitt}.
The global time, in particular cases, can be found from the Hamiltonian constraint
in homogeneous models, see {e.g.}~\cite{Kuchar,PIntern,Misner:1969hg}.
Charles W.~Misner introduced the intrinsic time as the logarithm of the hypersurface volume
in a homogeneous mixmaster universe~\cite{Misner:1969hg}.
The problem of time and the corresponding energy exists in cosmological models
because of absence of any outer observer watching the evolution of closed manifolds.
The conformal Dirac's mapping~\cite{Dirac:1958jc} allows to extract a local intrinsic
time which is a scalar under diffeomorphisms. In opposite to the asymptotically flat
spaces, Geometrodynamics of closed manifolds is related to non-conservative Hamiltonian 
systems.

\section{Many-fingered intrinsic time}

The spacetime ${\mathcal M}={\mathbb{R}}^1\times\Sigma_t$ with the metric tensor field
\begin{equation} \nonumber
{\bf g}:=g_{\mu\nu}(t, {\bf x}){\rm\bf d}x^\mu\otimes {\rm\bf d}x^\nu
\end{equation}
can be foliated into a family of space-like (hyper)surfaces $\Sigma_t,$ labeled
by the time coordinate $t$ with only spatial coordinates on each slice $(x^1, x^2, x^3)$.
Then, the evolution of space in time can be described in a natural way.
The components of the metric tensor in the ADM form are
\be\label{gmunu}
(g_{\mu\nu})= \left( \begin{array}{cc}
-N^2+N_iN^i& N_i \\
N_j& \gamma_{ij}
\end{array}
\right).
\ee
Here and below the Latin indices run as $i,j=1,2,3$.
The scalar field $N$ and the 3-vector field ${\bf N}$ extend
the coordinate system out of $\Sigma_t$. The first quadratic form
\be\label{spacemetric}
{\bfitgamma}:=\gamma_{ik}(t, {\bf x}){\rm\bf d}x^i\otimes {\rm\bf d}x^k
\ee
defines the induced metric on every slice $\Sigma_t$.

Let us implement the conformal transformation
\begin{equation}\label{Psifactor}
\gamma_{ij}:=\phi^4\tilde\gamma_{ij},\qquad \phi^4:=\sqrt[3]{\frac{\gamma}{f}},
\end{equation}
where, in addition to the determinant $\gamma$, the background static metric determinant
$f$ has appeared~\cite{Pavlov:2017auw}
\be \nonumber
f:={\rm det} (f_{ij}).
\ee
The introduction of the background metric will allow us to consider
not only asymptotically flat spaces but also general closed manifolds.
In addition to the space metric (\ref{spacemetric}), the background metric of 
the tangent space, Lie-dragged along the coordinate time evolution vector, 
can be introduced
\be\label{spacemetricback}
{\bf f}:=f_{ik}({\bf x}){\rm\bf d}x^i\otimes {\rm\bf d}x^k.
\ee
The Minkowskian metric as a background one was used in~\cite{Gor} for description
of gravitational problems in asymptotically flat spacetimes.
We claim that the background metric has to be chosen from the physical point of view.
The mapping of the Riemannian space with metric $\bfitgamma$ (\ref{spacemetric})
to the background space with the metric ${\bf f}$ (\ref{spacemetricback}) 
should be bijective. For this reason we suggest to consider a closed manifold. 
The conformal metric~(\ref{Psifactor})
\be\label{gammatilde}
{\tilde\bfitgamma}:=\tilde\gamma_{ik}(t, {\bf x}){\rm\bf d}x^i\otimes {\rm\bf d}x^k
\ee
is a tensor field, i.e., it is transformed according to the tensor representation
of the group of diffeomorphisms. The scaling factor $(\gamma/f)$ is a scalar field,
i.e., it is invariant under the diffeomorphisms.
To the conformal variables
\begin{equation}\label{generalized}
{{\tilde\gamma_{ij}:=\frac{\gamma_{ij}}{\sqrt[3]{\gamma /f}},\qquad
\tilde\pi^{ij}:=\sqrt[3]{\frac{\gamma}{f}}\left(\pi^{ij}-\frac{1}{3}\pi\gamma^{ij}\right)}},
\end{equation}
we add the canonical pair: {a local intrinsic time} $D$ and a trace of momentum
density $\pi$:
\begin{equation}\label{DiracTpi}
D:=-\frac{2}{3}\ln\sqrt{\frac{\gamma}{f}},\qquad \pi=2K\sqrt\gamma .
\end{equation}
Formulae (\ref{generalized}) and (\ref{DiracTpi}) define
(the scaled Dirac's mapping)~\cite{Pavlov:2017auw}
\begin{equation}\label{generalizedD}
(\gamma_{ij}, \pi^{ij})\mapsto (D,\pi; \tilde\gamma_{ij}, \tilde\pi^{ij}).
\end{equation}
The intrinsic time interval $\delta D$ that is a scalar under diffeomorphisms,
but without adding an auxiliary metric, was implemented in the symplectic
1-form~\cite{Murchadha:2012zz}.

The Poisson brackets between new variables in the extended phase space $\Gamma_D$ are
\begin{eqnarray}
\{D(t,{\bf x}), \pi(t,{\bf x}')\}&=&-\delta ({\bf x}-{\bf x}'),
\label{PoissonD} \\
\{\tilde\gamma_{ij}(t,{\bf x}),\tilde\pi^{kl}(t,{\bf x}')\}&=&\tilde\delta_{ij}^{kl}\delta ({\bf x}-{\bf x}'),
\label{PoissonG} \\
\{\tilde\pi^{ij}(t,{\bf x}),\tilde\pi^{kl}(t,{\bf x}')\}&=&\frac{1}{3}(\tilde\gamma^{kl}\tilde\pi^{ij}-
\tilde\gamma^{ij}\tilde\pi^{kl}) \delta ({\bf x}-{\bf x}'), \label{PoissonPi}
\end{eqnarray}
where
\be \nonumber
\tilde\delta_{ij}^{kl}:=\delta_i^k\delta_j^l+\delta_i^l\delta_j^k-
\frac{1}{3}\tilde\gamma^{kl}\tilde\gamma_{ij}
\ee
is the conformal Kronecker symbol.

The Hamiltonian constraint in the new variables yields the Lichnerowicz--York
differential equation
\be\label{H}
\left(\tilde\Delta-\frac{1}{8}\tilde{R}\right)\phi+
\frac{1}{8}\tilde{\pi}_{ij}\tilde{\pi}^{ij}\phi^{-7}-\frac{1}{12}K^2\phi^5
+\frac{1}{8}\tilde{T}_{\bot\bot}\phi^5=0.
\ee
Here $\tilde\nabla_k$ is the conformal connection associated with the conformal
metric $\tilde\gamma_{ij}$; quantity $\tilde{R}$ is the conformal Ricci scalar
expressed through the standard Ricci scalar $R$:
\be\label{Rscalar}
R= \frac{1}{\phi^4}\tilde{R}-\frac{8}{\phi^5}\tilde\Delta\phi.
\ee
The matter density is transformed according to
\be\label{Ttilde}
\tilde{T}_{\bot\bot}:=\phi^8 T_{\bot\bot}.
\ee

\section{Hamiltonian reduction and deparametrization}

The momentum density $\pi$ enters into the conformal Hamiltonian constraint
quadratically~(\ref{H}) as usual for relativistic theories. So, with the plus sign,
it can be expressed from the constraint
\bea \label{pfound}
&& \pi [\tilde\pi^{ij},\tilde\gamma_{ij}; \phi; x) \\
&=& 4\sqrt{3}\sqrt\gamma
\biggl[
\frac{1}{\phi^{5}}\left(\tilde\Delta-\frac{1}{8}\tilde{R}\right)\phi+
\frac{1}{8\phi^{12}}\tilde\pi_{ij}\tilde\pi^{ij}+
\frac{1}{8}\tilde{T}_{\bot\bot}\biggr]^{1/2}.\nonumber
\eea
Substituting the extracted $\pi$ (\ref{pfound}) into the ADM action~\cite{Arnowitt},
we obtain the functional presimplectic 1-form with dilaton field $D$ playing
the role of a local time~\cite{PavlovTwo}
\be \label{omega1}
\omega^1 = \int\limits_{\Sigma_D}
d^3x\biggl[\tilde\pi^{ij}d\tilde\gamma_{ij}-
\pi[\tilde\pi^{ij},\tilde\gamma_{ij}; D; x)d D
- N^i{\mathcal H}_i \biggr],
\ee
where ${\mathcal H}_i$ are supermomenta.

Taking in the role of the Hamiltonian $H(x)$ the integral over
the hypersurface of $\pi (x)$
\be\label{Hampi}
H=\int\limits_{\Sigma_t}{\rm d}^3x \,\pi^{ij}(x)\gamma_{ij}(x),
\ee
one should find the canonically conjugated global time $T$.
Let us extract the zero mode out of the scalar field that is the square root of ratio
of the determinants of metric tensors $\sqrt{\gamma /f}$
\be\label{extract}
\sqrt{\frac{\gamma}{f}}(x)=<\sqrt{\frac{\gamma}{f}}>+
\overline{\sqrt{\frac{\gamma}{f}}}(x),
\ee
where the mean value of it over the hypersurface $\Sigma_t$ is
\be\label{mean}
<\sqrt{\frac{\gamma}{f}}>:=
\frac{\int_{\Sigma_t}\,{\rm d}^3 y\sqrt{\gamma} (y)\sqrt{\gamma /f}(y)}
{\int_{\Sigma_t}\,{\rm d}^3 y\sqrt{\gamma}(y)}.
\ee
The second term in (\ref{extract}) is its residue with a zero mean value
\be
\int\limits_{\Sigma_t}\,{\rm d}^3 y\sqrt{\gamma} (y)
\overline{\sqrt{\frac{\gamma}{f}}}(y)=0.
\ee
The Poisson bracket between the mean value (\ref{mean}) and
the Hamiltonian~(\ref{Hampi}) reads
\be
\{<\sqrt{\frac{\gamma}{f}}>[\gamma_{ij}], H[\gamma_{ij},\pi^{ij}]\}=
\frac{3}{2}<\sqrt{\frac{\gamma}{f}}>.
\ee
Now we can define the global time
\be\label{global}
T(t):=-\frac{2}{3}\ln <\sqrt{\frac{\gamma}{f}}>
\ee
as the logarithmic function of the mean value over a hypersurface $\Sigma_t$
at every instant $t$.
Thus, one gets the canonical pair (\ref{Hampi}) and (\ref{global}) that commutes
to minus unit:
\be \label{TH}
\{T, H\}=-1.
\ee

The application of the perfect cosmological principle to formula~(\ref{global})
for the Friedmann--Robertson--Walker interval yields the ratio of Universe radius
$a(t)$ to the present day one $a_0$ from the standard cosmological conception
\be\nonumber
\frac{a(t)}{a_0}=e^{-(3/2) T}.
\ee
Here, for the background space a sphere of the present day radius $a_0$ was taken.
Thus, the global time is related to the observable redshift
\be \nonumber
z(t)=\frac{a_0-a(t)}{a(t)}.
\ee
The constant mean curvature slicing allows to obtain the global time by integration over
the manifold $\Sigma_t$
\be
\int\limits_{\Sigma_t}\,{\rm d}^3x\,\pi\frac{{\rm d}D}{{\rm d}t}=
H\frac{{\rm d}T}{{\rm d}t}. \nonumber
\ee

Then, we extract the zero mode of the field $\pi(x)$, that is a scalar density
\be\label{extractpi}
\pi (x)=\sqrt\gamma (x)<\pi>+\bar{\pi}(x),
\ee
where the mean value of $\pi(x)$ over a hypersurface $\Sigma_t$ is
\be\label{meanpi}
<\pi>:=\frac{\int_{\Sigma_t}\,{\rm d}^3 y \pi (y)}{\int_{\Sigma_t}\,
{\rm d}^3 y \sqrt{\gamma} (y)}.
\ee
Two third of the average of $\pi$ over $\Sigma_t$ is the global York time
and the conjugated variable (the Hamiltonian) is the volume of the hypersurface
in~\cite{Mercati}. Our approach is different because of these geometric characteristics 
possess in our case quite another essence. 
In general, canonical momenta are not defined within the hypersurface. 
They refer to motion in time of the original $\Sigma_t$. Intrinsic time is 
the variable constructed entirely out of the metric of the hypersurface.
In a sense, the roles of the Hamiltonian and global time are interchanged 
in~\cite{Mercati} in contradiction with the general principles.

The second term in (\ref{extractpi}) is the residue with zero mean value over 
a hypersurface $\Sigma_t$
\be
\int\limits_{\Sigma_t}\, {\rm d}^3 x\bar\pi (x)=0.
\ee
Thus, the mapping of phase space $\Gamma_D$ to the phase space $\bar\Gamma$ after
extraction of the global variables $T$ and $H$ is executed:
\be\nonumber
(D,\pi ;\tilde\gamma_{ij},\tilde\pi^{ij})\mapsto
(T, H,\bar\pi,\overline{\sqrt{\gamma /f}} ;\tilde\gamma_{ij},\tilde\pi^{ij}).
\ee

The Poisson brackets of the residues with the Hamiltonian are
\bea
\{\overline{\sqrt{\frac{\gamma}{f}}},H\}&=&
\frac{3}{2}\left(
e^{-(3/2) T}+\overline{\sqrt{\frac{\gamma}{f}}}\right),\label{PBr1}\\
\{\overline{\pi},H\}&=&0.\label{PBr2}
\eea
The residues commute with the global time
\be\nonumber
\{\overline{\sqrt{\gamma /f}}, T\}=0,
\ee
other commutation relations
$\{\bar{\pi},\bar{T}\},$ $\{\overline{\sqrt{\gamma /f}}, \pi\}$ are not required
further. Then, the integral Hamiltonian reduction to the phase space
$\widetilde\Gamma_D$ and the deparametrization procedure are performed
\be \nonumber
(D,\pi ;\tilde\gamma_{ij},\tilde\pi^{ij})
\mapsto (\tilde\gamma_{ij},\tilde\pi^{ij}).
\ee

That yields the action
\bea
S &=& \int\limits_{T_I}^{T_0} {\rm d}T\int\limits_{\Sigma_t} {\rm d}^3x\left[
\tilde\pi^{ij}\frac{{\rm d}\tilde\gamma_{ij}}{{\rm d} T}\right]
-\int\limits_{T_I}^{T_0} H\, {\rm d} T-\label{actionreduction}\\
&& \int\limits_{{\rm T}_I}^{{\rm T}_0}\, {\rm d}T\int\limits_{\Sigma_t} {\rm d}^3x N^i
{\mathcal H}_i\label{actiondiff}
\eea
with the Hamiltonian depending on the global time $T$
\be\label{reducedHam}
H[T(t),\overline{\sqrt{\gamma /f}}({\bf x});\tilde\pi^{ij}(x),\tilde\gamma_{ij}(x)]:=
\int\limits_{\Sigma_t}\,{\rm d}^3x\, {\mathcal H} (x)
\ee
with the Hamiltonian density
\bea \label{Hamdensity}
&& {\mathcal H} (x)\\
&=&4\sqrt{3}\sqrt\gamma
\biggl[
\frac{1}{\phi^{5}}\left(\tilde\Delta-\frac{1}{8}\tilde{R}\right)\phi+
\frac{1}{8\phi^{12}}\tilde\pi_{ij}\tilde\pi^{ij}+
\frac{1}{8}\tilde{T}_{\bot\bot}\biggr]^{1/2}.\nonumber
\eea

According to the commutation relations (\ref{PBr1}), (\ref{PBr2}), and
the Hamiltonian~(\ref{reducedHam}), the residue function
$\overline{\sqrt{\gamma /f}}({\bf x})$ does not depend on time $T$.
Functions of $\phi (x)$ and $\sqrt\gamma$ in Eq.~(\ref{Hamdensity}) can be expressed
via the global time and the residues, e.g., function $\phi^{12} (x)$ is a sum
of the time-dependent function and the space-dependent one:
\be
\phi^{12}(x)=\exp\left(-\frac{3}{2}T\right)+\overline{\sqrt{\frac{\gamma}{f}}}({\bf x}).
\ee

\section{Equations of motion}

The momentum constraints generate spatial diffeomorphisms by the last term
in action~(\ref{actiondiff}). Below, studying dynamics, we are not interested
in changes of coordinates of the hypersurface.
The energy of the universe is not conserved, it exponentially increases in time $T$.
The Hamiltonian flow is governed by Hamiltonian~(\ref{reducedHam})
with the Poisson brackets~(\ref{PoissonG}) and~(\ref{PoissonPi}):
\be
\frac{{\rm d}}{{\rm d}T}{\tilde\gamma}_{ij}({x})=
\int\limits_{\Sigma_t}\,{\rm d}^3{x}'
\{\tilde{\gamma}_{ij}({x}),{\tilde\pi}^{kl}({x}')\}
\frac{\delta}{\delta\tilde{\pi}^{kl}({x}')} H,\label{dgamma}
\ee
\bea \label{dpi}
\frac{{\rm d}}{{\rm d}T}{\tilde\pi}^{ij}({x})
&=& \int\limits_{\Sigma_t}\,{\rm d}^3{x}'
\{\tilde{\pi}^{ij}({x}),{\tilde\pi}^{kl}({x}')\}
\frac{\delta}{\delta\tilde{\pi}^{kl}({x}')} H
\nonumber \\
&+&\int\limits_{\Sigma_t}\,{\rm d}^3{x}'
\{\tilde{\pi}^{ij}({x}),{\tilde\gamma}_{kl}({x}')\}
\frac{\delta}{\delta\tilde{\gamma}_{kl}({x}')} H.
\eea
The functional derivative with respect to the momentum density reads
\be\label{varHpi}
\frac{\delta}{\delta\tilde{\pi}^{kl}({x}')} H [\phi;\tilde\pi^{ij},\tilde\gamma_{ij}]
=\frac{6\gamma (x')}{\phi^{12}(x'){\mathcal H}[\phi;\tilde\pi^{ij},\tilde\gamma_{ij};x')}
\tilde\pi_{kl}(x').
\ee
The derivative of the conformal metric with respect to the global time~(\ref{dgamma})
after application of~(\ref{PoissonG}) and~(\ref{varHpi}) becomes
\be \label{dgammadT}
\frac{{\rm d}}{{\rm d}T}\tilde\gamma_{ij}(x)
= \frac{12\gamma (x)\tilde\pi_{ij}(x)}
{\phi^{12}(x){\mathcal H}[\phi; \tilde\pi^{ij},\tilde\gamma_{ij};x)}.
\ee
Thus, the relation between the derivative of the generalized coordinates
$\tilde\gamma_{ij}$ with respect to the global time $T$ with the conjugate
generalized momenta $\tilde\pi^{ij}$ is obtained.

The first term in~(\ref{dpi}) is calculated easily
\bea \label{pipi}
&& \int\limits_{\Sigma_t}\,{\rm d}^3{x}'
\{\tilde{\pi}^{ij}({x}),{\tilde\pi}^{kl}({x}')\}
\frac{\delta}{\delta\tilde{\pi}^{kl}({x}')} H
\nonumber \\ && \quad
= -\frac{2\gamma (x)\tilde\gamma^{ij}(x)\tilde\pi^{kl}(x)\tilde\pi_{kl}(x)}
{\phi^{12}(x){\mathcal H}[\phi; \tilde\pi^{ij},\tilde\gamma_{ij};x)}.
\eea

One can write the functional derivative of the Hamiltonian density $\mathcal{H}$
with respect to the conformal metric components $\tilde\gamma_{kl}$ as
\bea \nonumber
&& \frac{\delta}{\delta\tilde\gamma_{kl}(x)}{H}[\phi;\tilde\gamma_{ij},\tilde\pi^{ij}]
=\int\limits_{\Sigma_t}\,{\rm d}^3{y}\biggl(
-\frac{3\gamma(y)}{\phi^4(y){\mathcal H}(y)}
\\ &&
\times \frac{\delta}{\delta\tilde\gamma_{kl}(x)}\tilde{R}[\tilde\gamma_{ij},y)
+\frac{24\gamma(y)}{\phi^5(y){\mathcal H}(y)}
\frac{\delta}{\delta\tilde\gamma_{kl}(x)}\tilde\Delta_{y}\phi(y)\biggr). \nonumber
\eea
The functional derivative of the Ricci scalar with respect to the metric
coefficients reads
\bea
&& \frac{\delta}{\delta\tilde\gamma_{kl}(x)}R[\tilde\gamma_{ij};y)=\biggl(
-\tilde{R}^{kl}[\tilde\gamma_{kl};y)+\tilde\gamma^{kl}(y)
\tilde\Delta_{y}
\nonumber \\ && \quad
- \tilde\nabla_{y}^k\tilde\nabla_{y}^l\biggr)\delta(x-y).\nonumber
\eea
By summing up these four terms we obtain the functional derivative
\bea
&& \frac{\delta}{\delta\tilde\gamma_{kl}(x)}H[\phi,\tilde\gamma_{ij},\tilde\pi^{ij}]=
\frac{3\gamma (x)}{\phi^4(x){\mathcal H}(x)}\tilde{R}^{kl}
\nonumber \\ && \quad
+ 3\left(\tilde\nabla^k_x\tilde\nabla^l_x-\tilde\gamma^{kl}(x)\tilde\Delta_x\right)
\left(\frac{\gamma (x)}{\phi^4(x){\mathcal H}(x)}\right) \label{foursum}
\\&& \quad
+ 12(2\tilde\gamma^{km}\tilde\gamma^{ln}-\tilde\gamma^{kl}\tilde\gamma^{mn})\tilde\nabla_m
\left(\frac{\gamma (x)}{\phi^5(x){\mathcal H}(x)}\right)\tilde\nabla_n\phi (x).\nonumber
\eea

Finally, the derivative of the conformal momentum density with respect to
the global time~(\ref{dpi}) with application of the commutation relations between
conformal phase variables, taking into account~(\ref{pipi}) and~(\ref{foursum}),
becomes
\bea \nonumber
&& \frac{\rm d}{{\rm d}T}\tilde\pi^{ij}(x)
= - \frac{6\gamma (x)}{\phi^4(x){\mathcal H}(x)}\left(\tilde{R}^{ij}
- \frac{1}{6}\tilde\gamma^{ij}\tilde{R}\right)
\nonumber \\ &&
- \frac{2\gamma (x)}{\phi^{12}(x){\mathcal H}(x)}\tilde\gamma^{ij}(x)\tilde\pi^{kl}\tilde\pi_{kl}
\label{dpidT}\\ &&
- 3\biggl(\tilde\nabla^i\tilde\nabla^j
+ \tilde\nabla^j\tilde\nabla^i-\frac{4}{3}\tilde\gamma^{ij}
\tilde\nabla^k\tilde\nabla_k\biggr)
\left[\frac{\gamma (x)}{\phi^4 (x){\mathcal H}(x)}\right]
\nonumber\\ &&
- 24\left(\tilde\gamma^{ik}\tilde\gamma^{jl}
+ \tilde\gamma^{jk}\tilde\gamma^{il}
- \frac{2}{3}\tilde\gamma^{ij}\tilde\gamma^{kl}\right)\tilde\nabla_l\phi\tilde\nabla_k
\left[\frac{\gamma (x)}{\phi^5 (x){\mathcal H}(x)}\right].\nonumber
\eea

\section{Conclusions and discussion}

We demonstrated that in Geometrodynamics of closed manifolds,
it is possible to generalize the Misner approach and introduce the global time.
After application of the Hamiltonian reduction procedure,
one yields differential evolution equations for conformal
metric components and conformal momentum densities.
In opposite to the case of asymptotically flat, for 
closed manifolds we got non-conservative Hamiltonian systems.

The deviation of the mean value of the global times appears as a classical 
scalar field, it deserves additional attention to be physically interpreted. 
Note that it emerged without any modification of the Einstein's theory.
The Wheeler's thin sandwich conjecture in General Relativity under positive
lapse and some restriction is valid~\cite{Bartnik}. Thus, the lapse function
can be found from the Hamiltonian constraint. It was supposed that the
deviations from the mean value of the global time can play the role of static
gravitational potentials~\cite{Arbuzov:2010fz}.

Earlier to construct a scalar, the Minkowskian metric as a background one was used 
for asymptotically flat spaces~\cite{Gor}. The intrinsic time interval $\delta D$ 
as a scalar was implemented in the symplectic 1-form~\cite{Murchadha:2012zz}. 
For splitting one degree of freedom the average of the trace of the momentum 
density was used as York time in the shape dynamics~\cite{Mercati}. 
The key difference of our study is the consideration closed manifolds without 
the asymptotically flat space condition. Our choice is motivated by 
cosmological applications.

For interpretation of the latest data of the Hubble diagram, the global time
as the scale factor of the Friedmann--Robertson--Walker model was successfully
implemented in refs.~\cite{Zakharov:2010nf,PavlovMIPh}. The choice of conformal
variables allows to suggest a new interpretation of the redshift of distant
stellar objects. Both the changing volume of the Universe in standard
cosmology and the changing or masses of elementary particles
in conformal cosmology~\cite{Narlikar} can serve as the measure of time.
Above we have shown that the observed expansion of the Universe can be directly
related to the global time of the Universe.


\end{document}